# Structured Information in Metric Neural Networks

David Dominguez, Kostadin Koroutchev, Eduardo Serrano and Francisco B. Rodríguez

*EPS, Universidad Autónoma de Madrid, 28049 Madrid, Spain*

(Dated: July 29, 2005)

The retrieval abilities of spatially uniform attractor networks can be measured by the average overlap between patterns and neural states. We found that metric networks, with local connections, however, can carry information structured in blocks without any global overlap. There is a competition between global and blocks attractors. We propose a way to measure the block information, related to the fluctuation of the overlap. The phase-diagram with the transition from local to global information, shows that the stability of blocks grows with dilution, but decreases with the storage rate and disappears for random topologies.

## I. INTRODUCTION

Attractor neural networks (ANN) usually deal with global overlaps between patterns and neural states, employing uniform connectivity to perform the retrieval task. This is useful when the information is spatially distributed, because the patterns learning is robust against damaging pieces of the network. Nevertheless, starting from only local stimulus, no global information can be achieved. In many applications in pattern recognition, however, there is only information for blocks at disposal to the network, and one might define local overlaps. We study in this paper under which conditions, for the topology of the synaptic connectivity, local overlaps are stable or can help to retrieve full patterns. On the other hand, we propose a way to measure the block information, and compare with the global information.

More structured ANN than fully-connected architectures, have been recently studied, specially small-world topology [1],[2]. Such graphs, modelled by two parameters: the *connectivity* $\gamma \equiv K/N$ (the ratio of links degree $K$ per network size $N$), and the rate of random links $\omega$ (among all $K$ neighbors); can capture most facts of a wide range of networks [3],[4]. The load rate $\alpha = P/K$ (where $P$ is the number of learned patterns), and the overlap $m$ between neuron states and memorized patterns are the most used measures of the retrieval ability [5]. After some critical load $\alpha_c$ no retrieval is possible, and $m \to 0$. Alternatively, the mutual information alone, $MI(\alpha, m)$, is usefull to compare the performance of different topologies [6],[7].

If there is no global stimulus $m$, it holds $MI = 0$, and no global information is transmitted through the network. Local stimulus with finite $m$ usually dissipate for local networks (small $\omega$), but may propagate if the network is random-like (large $\omega$). On the other hand, the information about a pattern is invariant under the reverse transformation $m \to -m$, but it vanishes if only half of the neurons are flipped. Suppose, however, one flips blocks of pixels of a facial image, for instance one keeps the eyes dark but convert the black hair in white. The overall picture still is likely recognizable. This rises two questions: first, how to measure the information hidden in these blocks? second, which are the neural architectures able to convert this block's to global information? To answer to them is the main purpose of this work. Unlike previous works about bumps[8], we consider here the simplest model of binary uniform neurons $\sigma \in \pm 1$, with same metrical connectivity for all neurons, without any reinforcement mechanism, so we can single out the effect of topology on the structure of the retrieval attractor. Besides this issue, we study non-trivial block structures.

There are applications in control theory, as the synchronization of systems in robotic [9], as well as in cognitive neuroscience, where distint sensory organs receiving independent stimulus manage to give an overall response [10]. The standard uniformly distributed overlap $m$ along the network is the only source of information in an attractor network, if the topology is either largely connected or random. For metric topologies, where the neighbors are local, however, one might measure local overlaps $m_l$ inside blocks. A structured distribution of overlaps, even with a vanishing global $m$, can carry some spatially ordered information, if the overlap of the blocks oscillate between negative and positive $m_l \pm 1$. We call this a block information (B), to distinguish from the usual global retrieval (R). We propose in this paper a parameter to measure this block information, and show, by simulation and theory, the topological conditions $\gamma, \omega$ for the transition between phases B and R.

## II. THE TOPOLOGY AND INFORMATION

The synaptic couplings are $J_{ij} \equiv C_{ij} W_{ij}$, where $\mathbf{C} = \{C_{ij}\}$ is the topology matrix and in $\mathbf{W} = \{W_{ij}\}$ are the learning weights. The topology $\mathbf{C}$ splits in local and random links. The local links connect each neuron to its $K_l$ nearest neighbors, in a closed ring. The random links connect each neuron to $K_r$ others uniformly distributed along the network. Hence, the network degree is $K = K_l + K_r$. The network topology is then characterized by two parameters: the *connectivity* ratio, and the *randomness* ratio, defined respectively by:

$$\gamma = K/N, \ \omega = K_r/K, \qquad (1)$$

where $\omega$ plays the role of a rewiring probability in the *small-world* model (SW) [4].

The learning algorithm updates $\mathbf{W}$, according to the Hebb rule

$$W_{ij}^\mu = W_{ij}^{\mu-1} + \xi_i^\mu \xi_j^\mu. \qquad (2)$$

The network starts at $W_{ij}^0 = 0$, and after $P = \alpha K$ learning steps, it reaches a value $W_{ij} = \sum_\mu^p \xi_i^\mu \xi_j^\mu$. The learning stage is a slow dynamics, being stationary in the time scale of the faster retrieval stage, defined in the following. Note that the couplings $\mathbf{J}$ can be written as an adjacency list of neighbors, $\{i, j = i_k; k = 1...K\}$. So the storage cost of this network is $|\mathbf{J}| = N \times K$.

The network state at a given time $t$ is defined by a set of binary neurons, $\vec{\sigma}^t = \{\sigma_i^t \in \pm 1, i = 1,...,N\}$. Accordingly, each pattern $\vec{\xi}^\mu = \{\xi_i^\mu \in \pm 1, i = 1,...,N\}$, is a set of site-independent unbiased binary random variables, $p(\xi_i^\mu = \pm 1) = 1/2$. The network learns a set of independent patterns $\{\vec{\xi}^\mu, \mu = 1,...,P\}$. The task of the network is to retrieve a pattern (say, $\vec{\xi} \equiv \vec{\xi}^\mu$) starting from a neuron state $\vec{\sigma}^0$ which is close to it. This is achieved through the dynamics

$$\sigma_i^{t+1} = \text{sign}(h_i^t), \ h_i^t \equiv \sum_j J_{ij} \sigma_j^t, \ i = 1...N \qquad (3)$$

A stochastic macro-dynamics takes place due to the extensive learning of $P = \alpha K$ patterns.

For networks with a metric it is useful to define neighborhoods. Let the blocks $\lambda_l, l = 1...B$ be the sets $\{i = (l-1)L + k; k = 1...L\}$, of size $L = N/B$. The *block − overlap* between the neural states and the pattern restricted to the block $\lambda_l$ is:

$$m_l \equiv \frac{1}{L} \sum_{i \in \lambda_l} \xi_i \sigma_i, \qquad (4)$$

at an unspecified time step. We can define averages over blocks as: $\langle f_l \rangle_b \equiv \frac{1}{b} \sum_{l=1}^b f_l$.

The relevant order parameter are the mean *overlap* between the neural states and the pattern, $m_b$, and the blocks-variance $v_b$, given by

$$m_L \equiv \langle m_l \rangle_b, \ v_L \equiv \langle m_l^2 \rangle_b - m_L^2. \qquad (5)$$

The usual global *overlap* can also be written as $m_L = m_N \equiv \frac{1}{N} \sum_i \xi_i \sigma_i$. The standard *deviation* overlap is $\delta = \sqrt{v_L}$. If the size of the blocks are taken $L = 1$, then $b = N$ blocks can be considered pure noise, with $m_l = \pm 1$ If the global overlap is $m_{L=1} = 0$, then it holds $v_{L=1} = 1$, but there is no macroscopic order. On the other hand, if there is only one block $b = 1$, then $v_{L=N} = 0$. However, if the size is large but $1 \ll L \ll N$, the blocks carry local information.

Together with the overlap, one needs a measure of the load, given by $\alpha = |\{\vec{\xi}^\mu\}|/|\mathbf{J}| \equiv P/K$. In the limit $K, N \to \infty$, the overlap reads $m = \langle \sigma \xi \rangle_{\sigma,\xi}$. The brackets

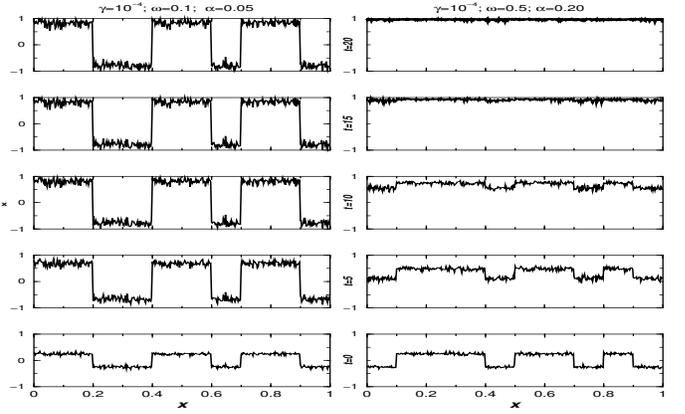

FIG. 1: Evolution of blocks with $m_l^0 = \pm 0.3$, from $t = 0$ (bottom) to $t = 20$ (top), for $\gamma = 10^{-4}$, with $N = 10^6$. Left: $\omega = 0.1, \alpha = 0.05$. Right: $\omega = 0.5, \alpha = 0.20$.

represent average over the joint distribution $p(\sigma, \xi)$, for a single neuron, understood as an ensemble distribution for the neuron states $\{\sigma_i\}$ and pattern $\{\xi_i\}$ [7]. With $p(\sigma, \xi)$ one can calculate the MI a quantity used to measure the knowledge that an observer at the output $\vec{\sigma}$) can get about the input $(\vec{\xi})$. It reads $MI[\sigma; \xi] = S[\sigma] - S[\sigma|\xi]$, where $S[\sigma] = 1[bit]$ and the conditional entropy is [7]:

$$S[\sigma|\xi] = -\frac{1+m}{2} \log_2 \frac{1+m}{2} - \frac{1-m}{2} \log_2 \frac{1-m}{2}. \qquad (6)$$

We define the global information rate as

$$i_m(\alpha, m_l) \equiv \alpha MI[\sigma; \xi], \qquad (7)$$

for independent neurons and patterns.

On the other hand, we consider a set of $b$ independent blocks of pattern overlaps, their distribution described by their mean $m_L$ and variance $v_L$. The information for blocks can be estimated from an output state made out of a signal term, $m_l$, with variance $v_L$, and a noise term, $m_i$, i.e.., the overlap of a single neuron, with variance $v_i = 1$. So the information is $MI[\vec{\sigma}, \{\vec{\xi}^\mu\}] \sim S[m_l + m_i] - S[m_i]$. For bimodal $m_l$, the block information rate is approximately

$$i_v(\alpha, m_l) = \alpha \log_2(1 + v_b). \qquad (8)$$

When the global (block) information increases, the block (global) information decreases.

### III. GLOBAL AND BLOCK PHASES

We simulated the dynamics in Eqs.(2-3) with the topology defined in Eq.(1), and observe the evolution in time for the block overlaps. It is illustrated in Fig.1, for a network of $N = 10^6$ neurons, with $\gamma = 10^{-4}$. The initial blocks are chosen at random with $m_l^0 = \pm 0.3$. In the left panel $\alpha = 0.05$ and $\omega = 0.1$, in the right panel it is $\alpha = 0.20$ and $\omega = 0.5$. The neuron overlaps $m_i$ are averaged over windows (of size $\delta L = 10^3$) inside the blocks $\lambda_l$ (of size $L = 10^5$), so the plotted curves are smoother than the actual $m_i$, but some structure is still seen inside $m_l$. While in the left panel, the blocks are retrieved as independent patterns, keeping their starting signals, but increasing the overlaps from $m_l^0 = \pm 0.3$ to $m_l^t \sim \pm 1.0$, in the right panel the blocks lose their starting signals, but the full pattern is completed.

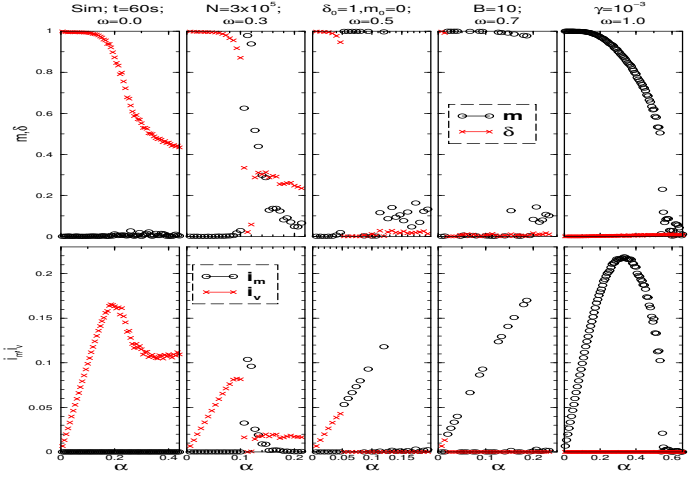

FIG. 2: Global and block overlaps $m, \delta$ (top) and informations $i_m, i_v$ (bottom), vs $\alpha$, for $b = 10$, $N = 3 \times 10^5$, $\gamma = 10^{-3}$, and $\omega$ from 0.0 (left) to 1.0 (right).

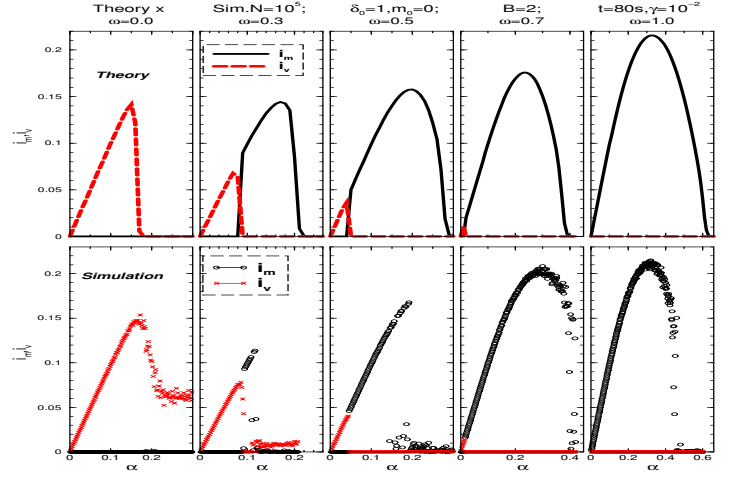

FIG. 4: Global (solid lines and circles, $i_m$) and block (dashed lines and $\times$, $i_v$) information, vs $\alpha$, for $b = 2$. Network with $\omega$ from 0 (left) to 1 (right). Top: theory. Bottom: simulation with $N = 10^5$, $\gamma = 10^{-2}$.

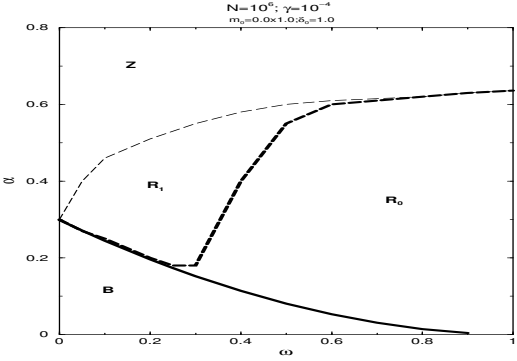

FIG. 3: Phase diagram ($\omega, \alpha$) for $N = 10^6$, $\gamma = 10^{-4}$ and $b = 2$. $B \equiv$ Block phase. $R_0(R_1) \equiv$ Retrieval with $m^0 = 0, \delta^0 = 1$. ($m^0 = 1, \delta^0 = 0$). $Z \equiv$ no information.

We studied the stationary states of the network, as a function of the topological parameters, $\omega$ and $\gamma$. A sample of the results for simulation is shown in Fig.2. The global and block overlaps, and the informations, are plotted for $\gamma = 10^{-3}$ and different randomness from $\omega = 0.0$ to $\omega = 1.0$. The neuron states start in a block structure with $b = 10$ and $m^0 = 0, \delta^0 = 1$. It can be seen that the maximal block (global) information decreases (increases) with $\omega$.

A reason for this behavior is that randomness decreases the mean-path-length between neurons, which facilitates the propagation of the information around the network. On the other way, locality increases the clustering of neurons, which stabilizes the formation of the blocks. If the connections are local, the information flows slowly over the network. thus the neurons can be eventually trapped in the blocks and the completion of the pattern is not allowed. It is worth to note that the network is able to retrieve the full pattern with $m^* \sim 1$ starting at $m^0 = 0$, thanks to the role of the $\pm$ block overlaps. This does not hold for initial condition $m^0 = 0, \delta^0 = 0$. Fig.2 also plots the information for the global and block states, $i_m, i_v$. The maximum of the block information, $i_v \sim 0.17$ for local topology, $\omega = 0.0$, is comparable to the maximum $i_m \sim 0.22$ for $\omega = 1.0$.

The network can exhibit two phases: a global *retrieval* (R) phase, with $m \neq 0, \delta_m = 0$, or a *block* deviation retrieval (B) phase, with $m = 0, \delta_m > 0$. When the network starts near a pattern, it will flow closer to that pattern if the load is lower than the saturation limit for the R phase, $\alpha_R(\omega)$. When the blocks of the network start successively near a pattern or its negative, it will flow closer to that blocks if the load is lower than the block saturation, $\alpha_B(\omega)$. For large $\omega$, the stable phase is the R, for small $\omega$, B is the stable phase. The phase diagram is shown in Fig.3, for a network of $N = 10^6$ neurons. We see that the transition from phase B to R holds at a larger $\alpha$ for more local networks than for more random networks.

Now we consider an extremely-diluted network. Let the neurons be distributed within $b$ blocks, for simplicity, each of size $L = N/b$, successively with positive and negative overlaps, $m_l = m_\pm$. Then the global overlap is $m = (m_+ + m_-)/2$ and the fluctuation between blocks is $\delta = (m_+ - m_-)/2$. The block overlaps can be written as $m_l = m + y_l \delta$, where $y_l \doteq \pm 1$ according to the block. An approximation for the local field of neurons at block $m_l$ at time step $t$ gives [5]

$$\xi h_l^t \equiv \omega m^t + (1-\omega)(m^t + y_l \delta^t)(1-\gamma b) + \Omega^t \qquad (9)$$

where $\xi$ is the pattern being retrieved. The pattern-interference noise is Gaussian distributed, $\Omega \doteq N(0, \Delta)$. Its variance $\Delta = \alpha r$ is given by the sum of random and local *feedback* terms, $r = \omega r_r + (1-\omega)r_l$, with $r_r = 1; r_l = (1+\chi)^2$. The *susceptibility* $\chi$ arises from the local connections. The correction term $1 - \gamma b$ accounts for the boundary effects between $m_\pm$ blocks.

Let it be $\gamma b \ll 1$. With the field in Eq.(9), the macrodynamics for the global and block overlaps are

$$\begin{aligned} m^{t+1} &= \langle \text{sign}(\xi h^t) \rangle_{y,z} \\ \delta^{t+1} &= \langle y \text{sign}(\xi h^t) \rangle_{y,z}, \end{aligned} \qquad (10)$$

with $\chi^t = \langle z \text{sign}(\xi h^t) \rangle_{y,z}/\sqrt{\alpha r_l}$, where the averages are over the distribution of $y$ and a Gaussian $z \doteq N(0, 1)$. We assume binary $y \in \pm 1$. There are two types of stationary states: (1) $m \neq 0, \delta = 0$, with $m = \text{erf}(m/\sqrt{2\alpha r_l})$ and (2) $m = 0, \delta > 0$, with $\delta = \text{erf}(\delta(1-\omega)/\sqrt{2\alpha r_l})$. The first is the usual Amit's solution[5]. The second is the block solution, which is stable if $(1-\omega) \geq \sqrt{\pi \alpha r/2}$. An adjust of the curve $B - R$ in the Fig.3 to $\omega \sim A\alpha^B$, gives $B = 0.51$, which fits well with this theoretical prediction, if one assumes $r \sim const.$ in the transition.

A comparison between theory and simulation is also given in Fig.4. The qualitative behavior of block retrieval with small $\omega$, and global retrieval with large $\omega$, as well as the transition for intermediate randomness at a given $\alpha(\omega)$ agree quite well, The maximal block and global information are also in agreement. Eq.(9) also explains why block retrieval fails for $L \sim K$, for which $\gamma b \sim 1$, and the boundaries between blocks are relevant compared to the bulk of the connected neurons. So, only diluted networks are able to stabilize blocks. However, the theoretical results start with $m^0 = 0.04$, and would never lead to $m^* \neq 0$ if $m^0 = 0$, as we observe in simulation. We believe that the finite size effect for $K < \infty$ in the simulation is the reason for this difference, and there is a lacks of theory to describe the $1/\sqrt{K}$ correction.

## IV. CONCLUSIONS

In this paper we have studied a new type of solution for an attractor neural network: the block overlap structure phase (B). Although resembling spurious states, where the network only recognize mixture of patterns, the B phase provides useful information, because the blocks are spatially ordered. The dependence stability of the B and the global retrieval overlap (R) phases with the topological parameters connectivity ($\gamma$) and randomness ($\omega$) was analyzed.

We found the transition from $R \to B$ takes place for $\alpha \leq \alpha_B \approx (1-\omega)^2$, and proposed a theory for strongly diluted networks which fits well with simulations. We also calculated the information, for both R and B phases.

The blocks behave as independent pieces of information: instead of the small number $P$ of patterns of size $N$ a diluted network can store, this phase is able to retrieve $b \times P$ patterns, each with size $N/b$. We believe that the existence of an information phase without any global overlap may play some relevant rule in natural neural networks, for instance, to manage a successful answer to stimuli activating separated cortical areas [10]. Also in many applications on pattern classification, such as image recognition, carrying local spatial information, the overlaps may have opposite signals in separate blocks, but an overall information could arise. Minor changes in the topology **C**, for instance suppressing symmetry constraints, lead to complex dynamics for the blocks, including cycles and chaos, which could model higher functions of the brain[11].

Supported by TIC01-572 and RyC/Spain grant.

Supported by TIC01-572 and RyC/Spain grant.

[1] C. Li and G. Chen, Phys. Rev. E **68**, 52901-4 (2003).
[2] P.N. McGraw and M. Menzinger Phys. Rev. E **68**, 047102-1 (2003).
[3] R. Albert and A. Barabasi, Rev. Mod. Phys. **74**, 47-97 (2002).
[4] D.J. Watts and S.H. Strogatz, Nature **393**, 440-442 (1998).
[5] J. Hertz, J. Krogh and, R. Palmer, *Introduction to the Theory of Neural Computation* (Addison-Wesley, Boston, 1991)
[6] M. Okada, Neural Network **9/8**, 1429-1458 (1996).
[7] D.R.C. Dominguez and D. Bolle, Phys. Rev. Lett **80**, 2961-2964 (1998).
[8] D. Dominguez, K. Koroutchev, E. Serrano and F. Rodriguez, LNCS **3173**, 14 (2004).
[9] S. Wermter, G. Palm, M. Elshaw, *Bio-Mimetic Neural Learning for Intelligent Robots* (Springer, Heidelberg, 2005).
[10] E. Rolls and A. Treves, *Neural Network and Brain Function* (Oxford University Press, Oxford, 2004).
[11] A. Damasio, *Descartes' Error: Emotion, Reason, and the Human Brain* (Grosset/Putnam, New York, 1994).

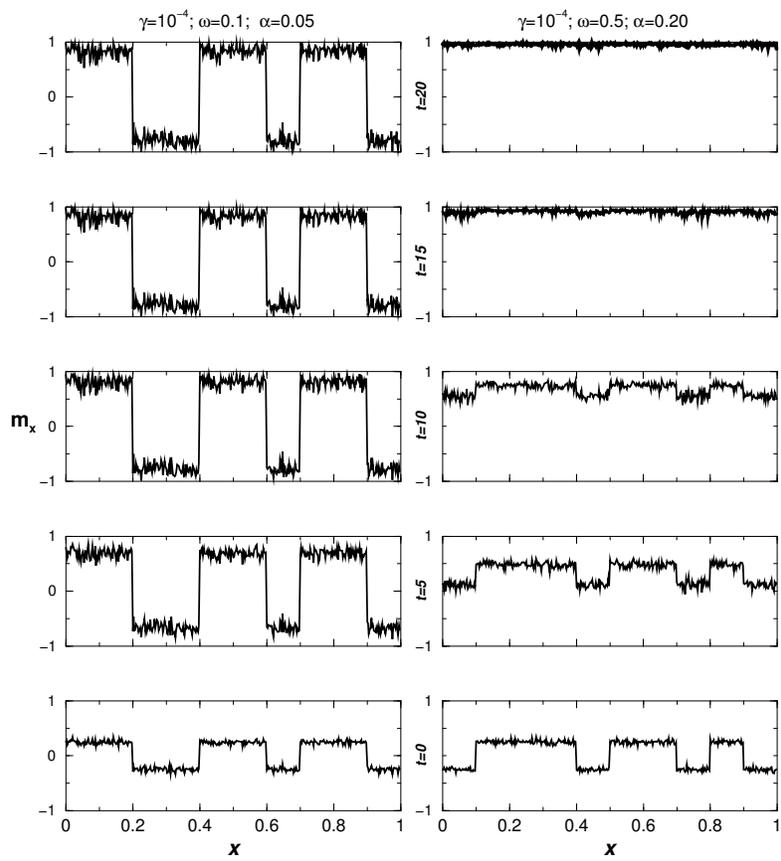

Figure 1

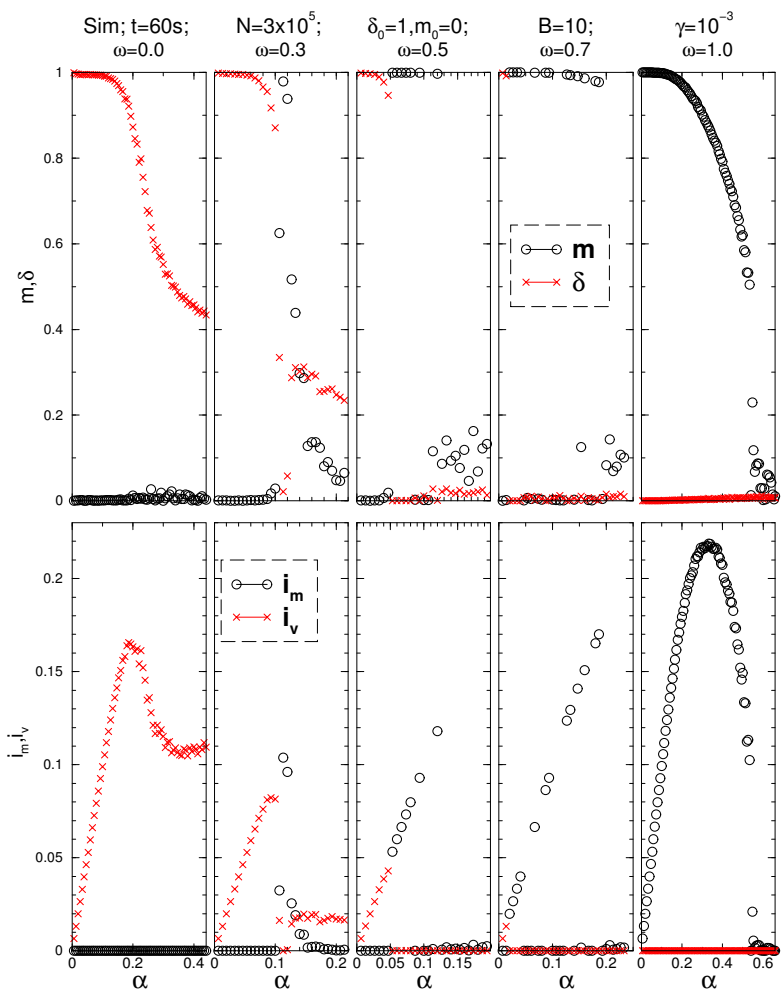

Figure 2

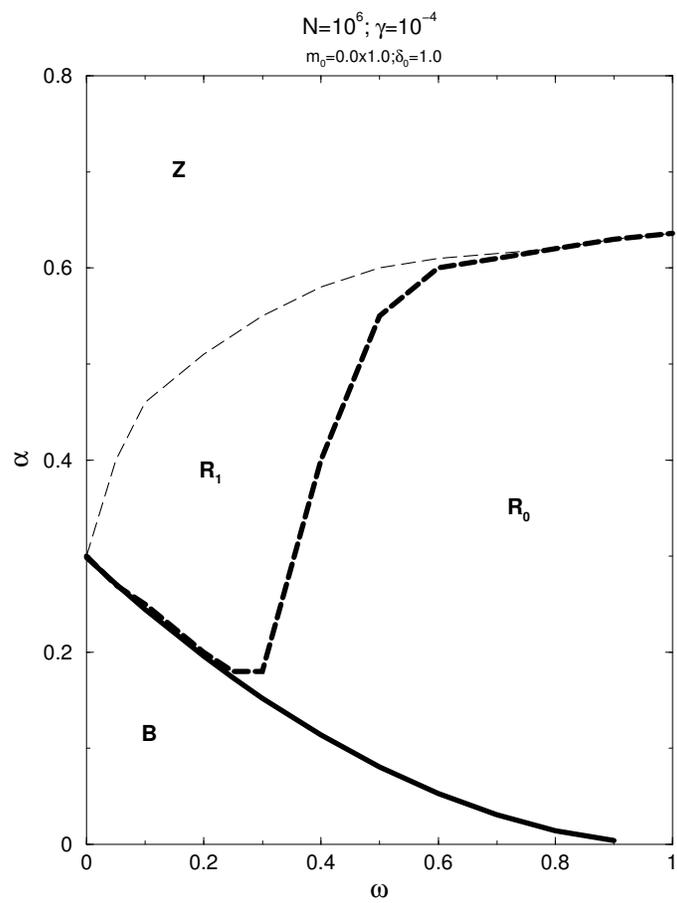

Figure 3

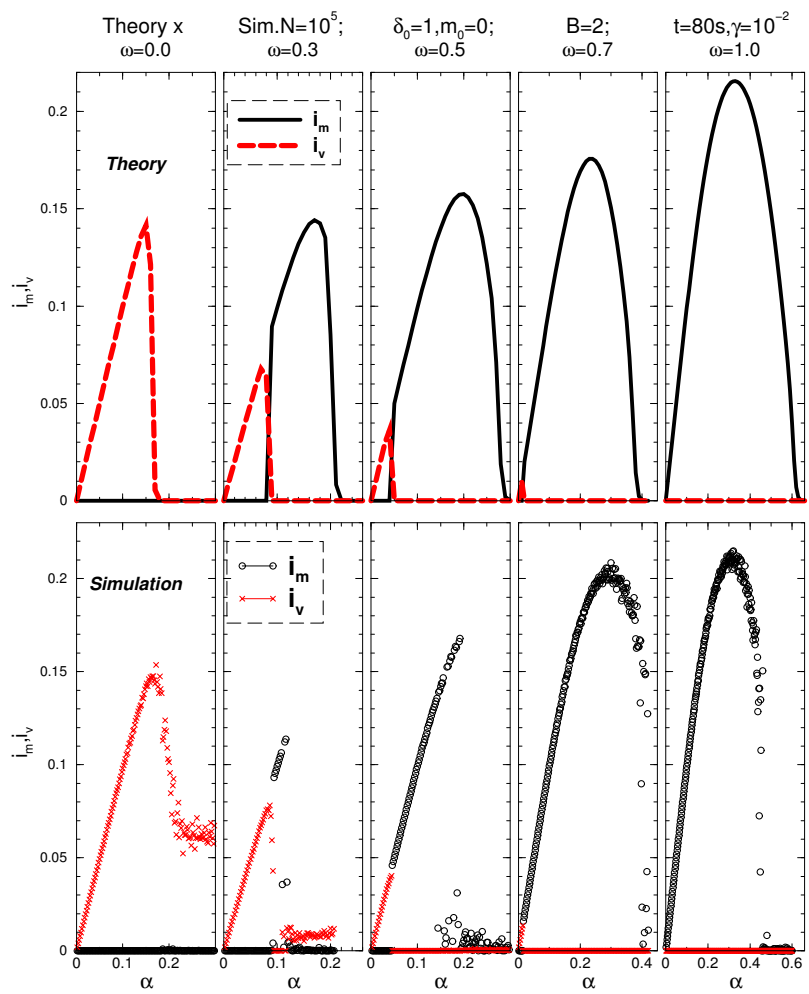

Figure 4